\documentclass[
    ,final            
  ]
  {aipproc}

\layoutstyle{6x9}

\newcommand{\acena}{\mbox{$\alpha$~Cen~A}}
\newcommand{\acenb}{\mbox{$\alpha$~Cen~B}}

\newcommand{\bhyi}{\mbox{$\beta$~Hyi}}
\newcommand{\bgem}{\mbox{$\beta$~Gem}}
\newcommand{\bvir}{\mbox{$\beta$~Vir}}
\newcommand{\nuind}{\mbox{$\nu$~Ind}}
\newcommand{\muara}{\mbox{$\mu$~Ara}}
\newcommand{\eboo}{\mbox{$\eta$~Boo}}

\newcommand{\Dnu}[1]{\Delta \nu_{#1}}
\newcommand{\muHz}{\mbox{$\mu$Hz}}
\def\Msol{\mbox{${M}_\odot$}}

\begin{document}

\title{Observing solar-like oscillations}



\author{Timothy R. Bedding}{
  address={School of Physics, University of Sydney, 2006, Australia} }

\author{Hans Kjeldsen}{
  address={Danish AsteroSeismology Centre (DASC), University of Aarhus,
Denmark} }

\begin{abstract}
The past few years have seen great progress in observing oscillations in
solar-type stars, lying on or just above the main sequence.  We review the
most recent results, most of which were obtained using high-precision
velocity measurements.  We also briefly discuss observations of more
evolved stars, namely G, K and M giants and supergiants.
\end{abstract}

\maketitle


\section{Main-sequence and subgiant stars}

The past few years have seen great progress in observing oscillations in
solar-type stars, lying on or just above the main sequence.  Most of the
recent results have come from high-precision Doppler measurements using
spectrographs such as CORALIE, HARPS, UCLES and UVES\@.  There are
ambitious plans to build SONG (Stellar Oscillations Network Group), which
will be a global network of small telescopes equipped with high-resolution
spectrographs and dedicated to asteroseismology and planet searches
\citep{GKF2006}.

Meanwhile, photometry from space is also making great prgress.  The WIRE
and MOST missions have reported oscillations in several stars, although not
without controversy, as discussed below.  The recent launch and successful
commissioning of COROT promises exciting new results (E. Michel et al.,
these Proceedings), as does the Kepler mission due for launch in early
2009.

Observations of solar-like oscillations are accumulating rapidly, and
measurement have now been reported for several main-sequence and subgiant
stars.  The following list includes the most recent observations and is
ordered according to decreasing stellar density (i.e., decreasing large
frequency separation):
\begin{description}

\item[$\tau$~Cet] (G8 V): this star was observed with HARPS by
T. C. Teixeira et al.\ (in prep.)

\item[70 Oph A] (K0 V): this is the main component of a spectroscopic visual
binary (the other component is K5~V).  It was observed over 6 nights with
HARPS by \citet{C+E2006}, who found $\Dnu{} = 162$\,\muHz{} but were not
able to give unambiguous mode identifications from these single-site data.

\item[\acenb] (K1~V): Oscillations in this star were detected from
single-site observations with CORALIE by \citet{C+B2003}.  Dual-site
observations with UVES and UCLES (see Fig.~\ref{fig.acenb}) allowed
measurement of nearly 40 modes and of the mode lifetime \citep{KBB2005}.

\begin{figure}
  \includegraphics[width=.8\textwidth]{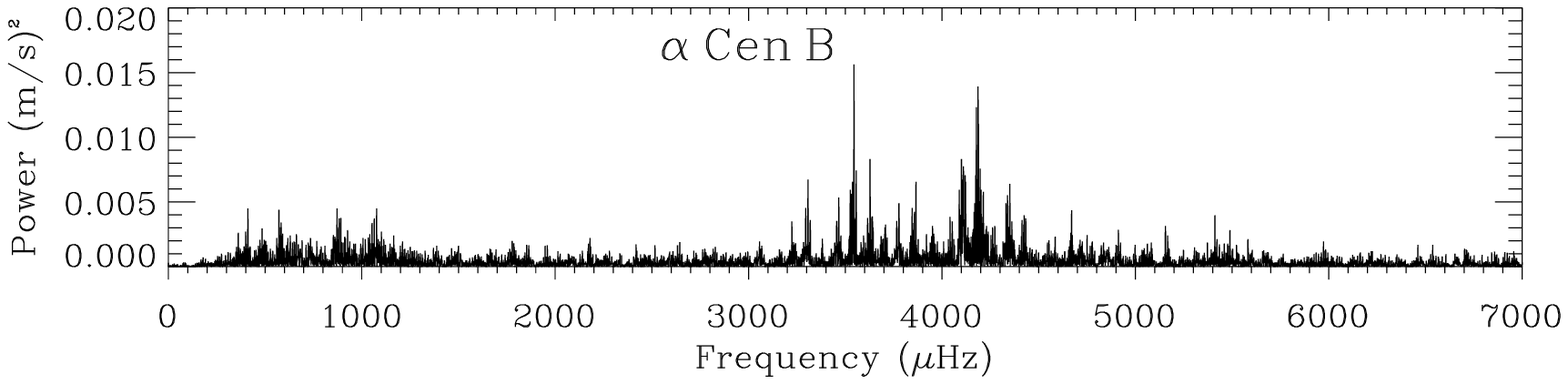}
  \caption{\label{fig.acenb} Power spectrum of \acenb{} from velocity
  observations.  Figure from \citet{KBB2005}. }
\end{figure}

\item[\acena] (G2~V): The clear detection of p-mode oscillations in
  \acena{} by \citet{B+C2002} using the CORALIE spectrograph represented a
  key moment in asteroseismology.  This was followed by a dual-site
  campaign on this star with UVES and UCLES \citep{BBK2004} that yielded
  more than 40 modes, with angular degrees of $l=0$ to~$3$ \citep{BKB2004}.
  The mode lifetime is about 2--4 days and there is now evidence of
  rotational splitting from photometry with the WIRE satellite analysed by
  \citet{FCE2006} (see Fig.~\ref{fig.fletcher}) and also from ground-based
  spectroscopy with HARPS \citep{BazBK2007}.

\begin{figure}
  \includegraphics[width=.8\textwidth]{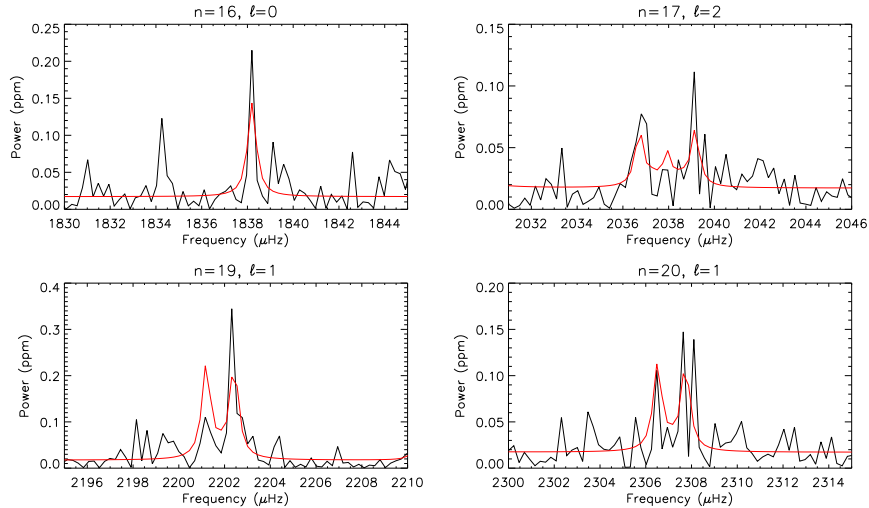}
  \caption{\label{fig.fletcher} Four oscillation modes in \acena{} from the
  WIRE power spectrum, with fits that indicate the linewidth and rotational
  splitting.  Figure from \citet{FCE2006}.}
\end{figure}

\item [\muara{}] (G3 V): this star has multiple planets.  Oscillations were
  measured over 8 nights using HARPS by \citet{BBS2005} (see
  Fig.~\ref{fig.muara}) and the results were modelled by \citet{BVB2005}.
  They found $\Dnu{} = 90$\,\muHz{} and identified over 40 frequencies,
  with possible evidence for rotational splitting.

\begin{figure}
  \includegraphics[width=.6\textwidth]{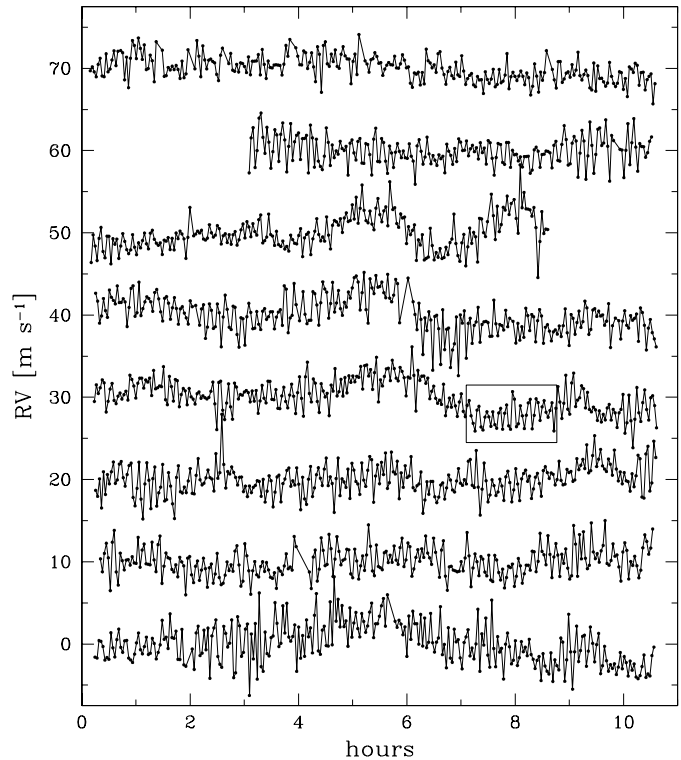}
  \caption{\label{fig.muara} Radial velocity time series of the star
  \muara{} made over 8 nights with the HARPS spectrograph.  Figure from
  \citet{BBS2005}.}
\end{figure}

\item[HD~49933] (F5~V): this is a primary target for the COROT space
mission and was observed over 10 nights with HARPS by \citet{MBC2005}.
They reported a surprisingly high level of velocity variability on
timescales of a few days.  This was also present as line-profile variations
and is therefore presumably due to stellar activity.  The observations
showed excess power from p-mode oscillations and the authors determined the
large separation ($\Dnu{} = 89\,\muHz$) but were not able to extract
individual frequencies.

\item[\bvir{}] (F9 V): oscillations in this star were detected in a
  weather-affected two-site campaign with ELODIE and FEROS by
  \citet{MLA2004b}.  Subsequently, \citet{CEDAl2005} used CORALIE with good
  weather but a single site, and reported 31 individual frequencies.  Those
  results were modelled by \citet{E+C2006}, who also reported tentative
  evidence for rotational splittings.  The large separation is 72\,\muHz.

\item[Procyon A] (F5 IV): At least eight separate velocity studies have
reported an excess in the power spectrum, beginning with that by
\citet{BGN91}.  The most recent examples were reported by \citet{MLA2004},
\citet{ECB2004}, \citet{BMM2004} and \citet{LKB2007}.  These studies agreed
on the location of the excess power (around 0.5--1.5\,mHz) but they
disagreed on the individual oscillation frequencies.  However, a consensus
has emerged that the large separation is about 55\,\muHz.

Controversy was generated when photometric observations obtained with the
MOST satellite by \citet{MKG2004} failed to reveal evidence for
oscillations.  However, \citet{BKB2005} argued that the MOST non-detection
was consistent with the ground-based data and \citet{R+RC2005} suggested
that the signature of oscillations is indeed present in the MOST data at a
low level.  Using space-based photometry with the WIRE satellite,
\citet{BKB2005b} extracted parameters for the stellar granulation and found
evidence for an excess due to p-mode oscillations.

A multi-site spectroscopic campaign on Procyon was carried out in January
2007, which has confirmed that this star does indeed show solar-like
oscillations (Arentoft et al. in prep.; Bedding et al., in prep.).

\item[\bhyi{}] (G2 IV): 
  In 2005 this star was the target for a two-site campaign using HARPS and
  UCLES \citet{BKA2007}, which resulted in the clear detection of mixed
  modes (see Fig.~\ref{fig.bhyi}). The large separation is 57.5\,\muHz.
  Combining this value with the angular diameter measured using
  interferometry gave the stellar mass to an accuracy of 2.7\%
  \citep{NDB2007}.  This is probably the most precise mass determination of
  a solar-type star that is not in a binary system, illustrating the power
  of combining asteroseismology and interferometry (see also
  \citealt{CMM2007} and references therein).

\begin{figure}
  \includegraphics[width=.6\textwidth]{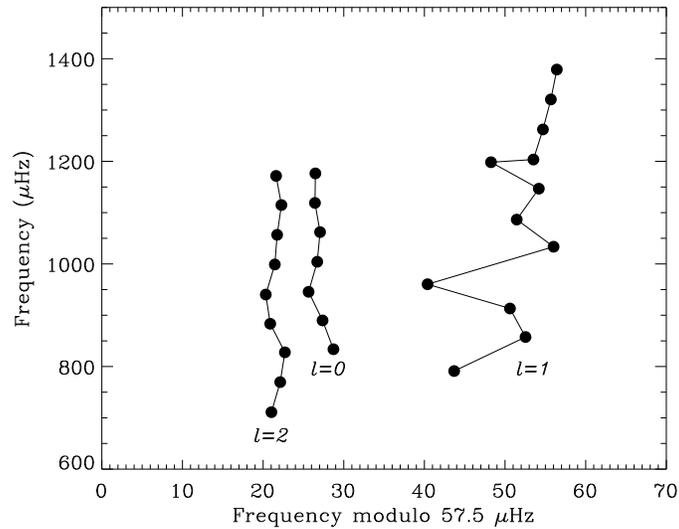} \caption{\label{fig.bhyi}
\'Echelle diagram of oscillation frequencies in \bhyi.  Figure from
\citet{BKA2007}.}
\end{figure}

\item[$\delta$~Eri] (K0 IV): \citet{CBE2003a} observed this star over 12
  nights in 2001 with CORALIE and found a large separation of 44\,\muHz.

\item[\eboo] (G0 IV): The claimed detection of oscillations by
\citet{KBV95}, based on fluctuations in Balmer-line equivalent-widths, has
now been confirmed by further equivalent-width and velocity measurements by
the same group \citep{KBB2003} and also by independent velocity
measurements with the CORALIE spectrograph \citep{CEB2005}.  With the
benefit of hindsight, we can now say that \eboo{} was the first star for
which the large separation and individual frequencies were measured.
However, there is still disagreement on some of the individual frequencies,
which reflects the subjective way in which genuine oscillation modes must
be chosen from noise peaks and corrected for daily aliases.  Fortunately,
the large separation is $\Dnu= 40$\,\muHz, which is half way between
integral multiples of the 11.57-\muHz{} daily splitting (40/11.57 = 3.5).
Even so, daily aliases are problematic, especially because some of the
modes in \eboo{} appear to be shifted by avoided crossings.

Spaced-based observations of \eboo, made with the MOST satellite, have
generated considerable controversy.  \citet{GKR2005} showed an amplitude
spectrum (their Fig.~1) that rises towards low frequencies in a fashion
that is typical of noise from instrumental and stellar sources.  However,
they assessed the significance of individual peaks by their strength
relative to a fixed horizontal threshold, which naturally led them to
assign high significance to peaks at low frequency.  They did find a few
peaks around 600\,\muHz{} that agreed with the ground-based data, but they
also identified eight of the many peaks at much lower frequency
(130--500\,\muHz), in the region of rising power, as being due to
low-overtone p-modes.  Those peaks do line up quite well with the regular
40\,\muHz{} spacing, but extreme caution is needed before these peaks are
accepted as genuine.  This is especially true given that the orbital
frequency of the spacecraft (164.3\,\muHz) is, by bad luck, close to four
times the large separation of \eboo{} (164.3/40 = 4.1).  Models of \eboo{}
based on the combination of MOST and ground-based frequencies have been
made by \citet{SDG2006}.

\item[\nuind] (G0 IV): this a metal-poor subgiant ($\mbox{[Fe/H]} = -1.4$)
  which was observed from two sites using UCLES and CORALIE
  \citep{BBC2006,CKB2007}.  The large separation of 24\,\muHz, combined
  with the position of the star in the H-R diagram, indicated that the star
  has a low mass a low mass ($0.85\pm0.04\,\Msol$) and is at least 9\,Gyr
  old (see Fig.~\ref{fig.nuind}).

\begin{figure}
  \includegraphics[width=.5\textwidth]{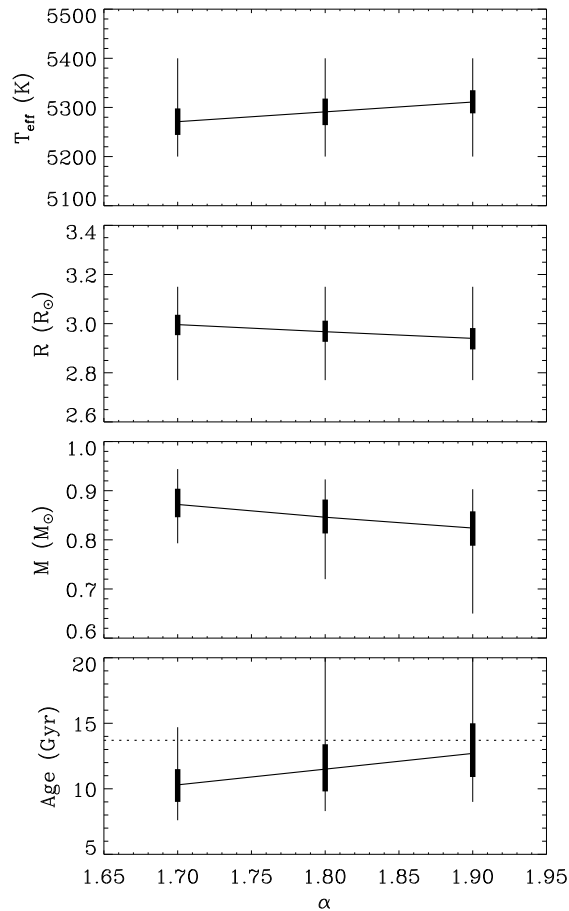}
  \caption{\label{fig.nuind} Parameters of \nuind{} for three different
  choices of $\alpha$, the mixing-length parameter.  The thin error bars
  show the range of each parameter based on classical measurements alone
  (luminosity and temperature), while the thick bars include the constraint
  provided by the large frequency separation.  The dashed line at 13.7 Gyr
  indicates the upper limit set by age of the universe from cosmology.
  Figure from \citet{BBC2006}.}
\end{figure}

\end{description}


\section{G and K Giants}

Oscillations in a few G and K giant stars have been detected using
ground-based velocity observations, indicating periods of a few to several
hours.  Oscillations in $\xi$~Hya (G7 III) \citep{FCA2002,SKB2004} measured
using CORALIE appear to have mode lifetimes of only about 2\,days.  The
CORALIE and ELODIE spectrographs have been used to find excess power and a
possible large separation for both $\epsilon$~Oph (G9 III)
\citep{DeRBC2006} and $\eta$~Ser (K0 III) \citep{BdRM2004}.
\citet{HADeR2006} have analysed the line-profile variations and found
evidence for non-radial oscillations in some of these stars.  Most
recently, \citet{H+Z2007} reported evidence for oscillations in the K giant
star \bgem, although this was based on a rather short time series (20
hours).

Spaced-based photometry of $\alpha$~UMa (K0~III) with WIRE produced
evidence for multi-mode oscillations \citep{BCL2000}.  More recently,
photometry of $\epsilon$~Oph with MOST confirmed the ground-based detection
of oscillations, establishing the large separation to be 5.1\,\muHz{} and
indicating a mode lifetime of 2--3\,d \citep{BMdeR2007}.  Mode lifetimes of
a few days are the same as those in the Sun but these giant stars oscillate
about 50 times more slowly.  Such short lifetimes, if confirmed as typical,
would significantly limit the prospects for asteroseismology on red giants.

Photometry of clusters is another way to search for oscillations in
K~giants.  \citet{E+G96} observed K~giants in the globular cluster 47~Tuc
using the {\em Hubble Space Telescope\/} over 38.5\,hr and found variables
with periods of 2--4 days and semi-amplitudes of 5--15\,mmag.  More
recently, multi-site ground-based observations have given tentative
evidence for oscillations in two open clusters, namely M67 \citep{SBK2007}
and M4 \citep{FBG2007}.

Finally, we note that photometry of Arcturus (K1.5 III) has been reported
by \citet{TCE2007} from 2.5\,yr of observations made by the Solar Mass
Ejection Imager (SMEI) on board the Coriolis satellite.  The observations
indicate a strong mode at 3.5\muHz{} with a damping time of 24\,d.  This
mode lifetime is also short when we take into account the period of the
oscillations (3.3\,d).

\section{M Giants and supergiants}

If we define solar-like oscillations to be those excited and damped by
convection then we might expect to see such oscillations in all stars on
the cool side of the instability strip.  Evidence for solar-like
oscillations in semiregular variables, based on visual observations by
groups such as the AAVSO, has already been reported. This was based on the
amplitude variability of these stars \citep{ChDKM2001} and on the
Lorenztian profiles of the power spectra \citep{BKK2005}.

\begin{figure}
  \includegraphics[width=.6\textwidth]{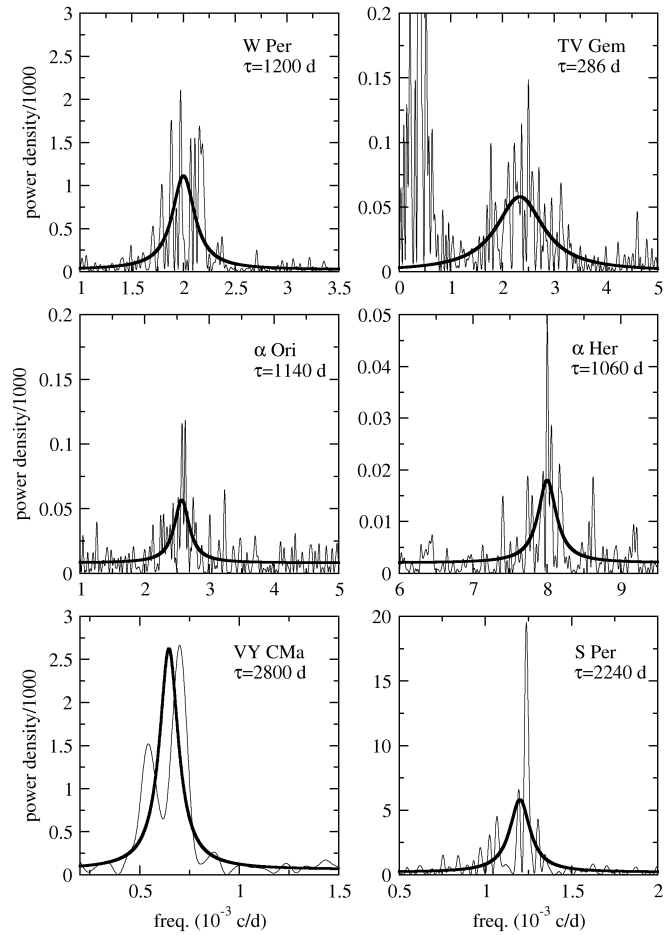} \caption{\label{fig.kiss}
  Power spectra of red supergiants from visual observations (thin lines)
  with Lorentzian fits (thick lines).  Figure from \citet{KSB2006}.}
\end{figure}

Recently, \citet{KSB2006} used visual observations from the AAVSO to show
that red supergiants such as Betelgeuse ($\alpha$~Ori), which have masses
of 10--30\Msol, also have Lorenztian profiles in their power spectra (see
Fig.~\ref{fig.kiss}).

\begin{theacknowledgments}
This work was supported financially by the Australian Research Council, the
Science Foundation for Physics at the University of Sydney, and the Danish
Natural Science Research Council.
\end{theacknowledgments}

\bibliographystyle{mn2e}

\begin{thebibliography}{}

\bibitem[\protect\citeauthoryear{{Barban}, {de Ridder}, {Mazumdar}
  et~al.,}{{Barban} et~al.}{2004}]{BdRM2004}
{Barban} C.,  {de Ridder} J.,  {Mazumdar} A.,    et~al., 2004, in Danesy D.,
  ed., SOHO 14/GONG 2004 Workshop, Helio- and Asteroseismology: Towards a
  Golden Future ESA SP-559.
p.~113

\bibitem[\protect\citeauthoryear{{Barban}, {Matthews}, {de Ridder}
  et~al.,}{{Barban} et~al.}{2007}]{BMdeR2007}
{Barban} C.,  {Matthews} J.~M.,  {de Ridder} J.,    et~al., 2007, A\&A, 468,
  1033

\bibitem[\protect\citeauthoryear{{Bazot}, {Bouchy}, Kjeldsen, Charpinet,
  Laymand \& {Vauclair}}{{Bazot} et~al.}{2007}]{BazBK2007}
{Bazot} M.,  {Bouchy} F.,  Kjeldsen H.,  Charpinet S.,  Laymand M.,
  {Vauclair} S.,  2007, A\&A, 470, 295

\bibitem[\protect\citeauthoryear{{Bazot}, {Vauclair}, {Bouchy} \&
  {Santos}}{{Bazot} et~al.}{2005}]{BVB2005}
{Bazot} M.,  {Vauclair} S.,  {Bouchy} F.,    {Santos} N.~C.,  2005, A\&A, 440,
  615

\bibitem[\protect\citeauthoryear{{Bedding}, {Butler}, {Carrier}
  et~al.,}{{Bedding} et~al.}{2006}]{BBC2006}
{Bedding} T.~R.,  {Butler} R.~P.,  {Carrier} F.,    et~al., 2006, ApJ, 647, 558

\bibitem[\protect\citeauthoryear{{Bedding}, {Kiss}, {Kjeldsen}
  et~al.,}{{Bedding} et~al.}{2005}]{BKK2005}
{Bedding} T.~R.,  {Kiss} L.~L.,  {Kjeldsen} H.,    et~al., 2005, MNRAS, 361,
  1375

\bibitem[\protect\citeauthoryear{{Bedding}, {Kjeldsen}, Arentoft
  et~al.,}{{Bedding} et~al.}{2007}]{BKA2007}
{Bedding} T.~R.,  {Kjeldsen} H.,  Arentoft T.,    et~al., 2007, ApJ, 663, 1315

\bibitem[\protect\citeauthoryear{{Bedding}, {Kjeldsen}, {Bouchy}
  et~al.,}{{Bedding} et~al.}{2005}]{BKB2005}
{Bedding} T.~R.,  {Kjeldsen} H.,  {Bouchy} F.,    et~al., 2005, A\&A, 432, L43

\bibitem[\protect\citeauthoryear{Bedding, Kjeldsen, Butler et~al.,}{Bedding
  et~al.}{2004}]{BKB2004}
Bedding T.~R.,  Kjeldsen H.,  Butler R.~P.,    et~al., 2004, ApJ, 614, 380

\bibitem[\protect\citeauthoryear{{Bouchy}, {Bazot}, {Santos}, {Vauclair} \&
  {Sosnowska}}{{Bouchy} et~al.}{2005}]{BBS2005}
{Bouchy} F.,  {Bazot} M.,  {Santos} N.~C.,  {Vauclair} S.,    {Sosnowska} D.,
  2005, A\&A, 440, 609

\bibitem[\protect\citeauthoryear{Bouchy \& {Carrier}}{Bouchy \&
  {Carrier}}{2002}]{B+C2002}
Bouchy F.,  {Carrier} F.,  2002, A\&A, 390, 205

\bibitem[\protect\citeauthoryear{Bouchy, Maeder, Mayor, M{\'e}gevand, Pepe \&
  Sosnowska}{Bouchy et~al.}{2004}]{BMM2004}
Bouchy F.,  Maeder A.,  Mayor M.,  M{\'e}gevand D.,  Pepe F.,    Sosnowska D.,
  2004, Nat, 432, 7015

\bibitem[\protect\citeauthoryear{Brown, Gilliland, Noyes \& Ramsey}{Brown
  et~al.}{1991}]{BGN91}
Brown T.~M.,  Gilliland R.~L.,  Noyes R.~W.,    Ramsey L.~W.,  1991, ApJ, 368,
  599

\bibitem[\protect\citeauthoryear{{Bruntt}, {Kjeldsen}, {Buzasi} \&
  {Bedding}}{{Bruntt} et~al.}{2005}]{BKB2005b}
{Bruntt} H.,  {Kjeldsen} H.,  {Buzasi} D.~L.,    {Bedding} T.~R.,  2005, ApJ,
  633, 440

\bibitem[\protect\citeauthoryear{Butler, Bedding, Kjeldsen et~al.,}{Butler
  et~al.}{2004}]{BBK2004}
Butler R.~P.,  Bedding T.~R.,  Kjeldsen H.,    et~al., 2004, ApJ, 600, L75

\bibitem[\protect\citeauthoryear{Buzasi, Catanzarite, Laher et~al.,}{Buzasi
  et~al.}{2000}]{BCL2000}
Buzasi D.~L.,  Catanzarite J.,  Laher R.,    et~al., 2000, ApJ, 532, L133

\bibitem[\protect\citeauthoryear{Carrier, Bouchy \& Eggenberger}{Carrier
  et~al.}{2003}]{CBE2003a}
Carrier F.,  Bouchy F.,    Eggenberger P.,  2003, in Thompson M.~J.,  Cunha
  M.~S.,   Monteiro M. J. P. F.~G.,  eds, Asteroseismology Across the HR
  Diagram Kluwer, p.~P311

\bibitem[\protect\citeauthoryear{Carrier \& {Bourban}}{Carrier \&
  {Bourban}}{2003}]{C+B2003}
Carrier F.,  {Bourban} G.,  2003, A\&A, 406, L23

\bibitem[\protect\citeauthoryear{{Carrier} \& {Eggenberger}}{{Carrier} \&
  {Eggenberger}}{2006}]{C+E2006}
{Carrier} F.,  {Eggenberger} P.,  2006, A\&A, 450, 695

\bibitem[\protect\citeauthoryear{{Carrier}, {Eggenberger} \&
  {Bouchy}}{{Carrier} et~al.}{2005}]{CEB2005}
{Carrier} F.,  {Eggenberger} P.,    {Bouchy} F.,  2005, A\&A, 434, 1085

\bibitem[\protect\citeauthoryear{{Carrier}, {Eggenberger}, {D'Alessandro} \&
  {Weber}}{{Carrier} et~al.}{2005}]{CEDAl2005}
{Carrier} F.,  {Eggenberger} P.,  {D'Alessandro} A.,    {Weber} L.,  2005,
  NewA, 10, 315

\bibitem[\protect\citeauthoryear{{Carrier}, {Kjeldsen}, {Bedding}
  et~al.,}{{Carrier} et~al.}{2007}]{CKB2007}
{Carrier} F.,  {Kjeldsen} H.,  {Bedding} T.~R.,    et~al., 2007, A\&A, 470,
  1059

\bibitem[\protect\citeauthoryear{Christensen-Dalsgaard, {Kjeldsen} \&
  {Mattei}}{Christensen-Dalsgaard et~al.}{2001}]{ChDKM2001}
Christensen-Dalsgaard J.,  {Kjeldsen} H.,    {Mattei} J.~A.,  2001, ApJ, 562,
  L141

\bibitem[\protect\citeauthoryear{{Creevey}, {Monteiro}, {Metcalfe}
  et~al.,}{{Creevey} et~al.}{2007}]{CMM2007}
{Creevey} O.~L.,  {Monteiro} M.~J.~P.~F.~G.,  {Metcalfe} T.~S.,    et~al.,
  2007, ApJ, 659, 616

\bibitem[\protect\citeauthoryear{{De Ridder}, {Barban}, {Carrier} et~al.,}{{De
  Ridder} et~al.}{2006}]{DeRBC2006}
{De Ridder} J.,  {Barban} C.,  {Carrier} F.,    et~al., 2006, A\&A, 448, 689

\bibitem[\protect\citeauthoryear{Edmonds \& Gilliland}{Edmonds \&
  Gilliland}{1996}]{E+G96}
Edmonds P.~D.,  Gilliland R.~L.,  1996, ApJ, 464, L157

\bibitem[\protect\citeauthoryear{{Eggenberger} \& {Carrier}}{{Eggenberger} \&
  {Carrier}}{2006}]{E+C2006}
{Eggenberger} P.,  {Carrier} F.,  2006, A\&A, 449, 293

\bibitem[\protect\citeauthoryear{Eggenberger, {Carrier}, {Bouchy} \&
  {Blecha}}{Eggenberger et~al.}{2004}]{ECB2004}
Eggenberger P.,  {Carrier} F.,  {Bouchy} F.,    {Blecha} A.,  2004, A\&A, 422,
  247

\bibitem[\protect\citeauthoryear{{Fletcher}, {Chaplin}, {Elsworth}, {Schou} \&
  {Buzasi}}{{Fletcher} et~al.}{2006}]{FCE2006}
{Fletcher} S.~T.,  {Chaplin} W.~J.,  {Elsworth} Y.,  {Schou} J.,    {Buzasi}
  D.,  2006, MNRAS, 371, 935

\bibitem[\protect\citeauthoryear{Frandsen, {Bruntt}, {Grundahl}
  et~al.,}{Frandsen et~al.}{2007}]{FBG2007}
Frandsen S.,  {Bruntt} H.,  {Grundahl} F.,    et~al., 2007, A\&A, in press

\bibitem[\protect\citeauthoryear{Frandsen, {Carrier}, {Aerts} et~al.,}{Frandsen
  et~al.}{2002}]{FCA2002}
Frandsen S.,  {Carrier} F.,  {Aerts} C.,    et~al., 2002, A\&A, 394, L5

\bibitem[\protect\citeauthoryear{{Grundahl}, {Kjeldsen}, {Frandsen}
  et~al.,}{{Grundahl} et~al.}{2006}]{GKF2006}
{Grundahl} F.,  {Kjeldsen} H.,  {Frandsen} S.,    et~al., 2006, Mem. Soc.
  Astron. Ital., 77, 458

\bibitem[\protect\citeauthoryear{{Guenther}, {Kallinger}, {Reegen}
  et~al.,}{{Guenther} et~al.}{2005}]{GKR2005}
{Guenther} D.~B.,  {Kallinger} T.,  {Reegen} P.,    et~al., 2005, ApJ, 635, 547

\bibitem[\protect\citeauthoryear{Hatzes \& Zechmeister}{Hatzes \&
  Zechmeister}{2007}]{H+Z2007}
Hatzes A.~P.,  Zechmeister M.,  2007, Astrophys. J., Lett., in press,
  {\tt\small arXiv:astro-ph/0709.1406}

\bibitem[\protect\citeauthoryear{{Hekker}, {Aerts}, {De Ridder} \&
  {Carrier}}{{Hekker} et~al.}{2006}]{HADeR2006}
{Hekker} S.,  {Aerts} C.,  {De Ridder} J.,    {Carrier} F.,  2006, A\&A, 458,
  931

\bibitem[\protect\citeauthoryear{{Kiss}, {Szabo} \& {Bedding}}{{Kiss}
  et~al.}{2006}]{KSB2006}
{Kiss} L.~L.,  {Szabo} G.~M.,    {Bedding} T.~R.,  2006, MNRAS, 372, 1721

\bibitem[\protect\citeauthoryear{Kjeldsen, Bedding, Baldry et~al.,}{Kjeldsen
  et~al.}{2003}]{KBB2003}
Kjeldsen H.,  Bedding T.~R.,  Baldry I.~K.,    et~al., 2003, AJ, 126, 1483

\bibitem[\protect\citeauthoryear{Kjeldsen, Bedding, Butler et~al.,}{Kjeldsen
  et~al.}{2005}]{KBB2005}
Kjeldsen H.,  Bedding T.~R.,  Butler R.~P.,    et~al., 2005, ApJ, 635, 1281

\bibitem[\protect\citeauthoryear{Kjeldsen, Bedding, Viskum \&
  Frandsen}{Kjeldsen et~al.}{1995}]{KBV95}
Kjeldsen H.,  Bedding T.~R.,  Viskum M.,    Frandsen S.,  1995, AJ, 109, 1313

\bibitem[\protect\citeauthoryear{{Leccia}, {Kjeldsen}, {Bonanno}, {Claudi},
  {Ventura} \& {Patern\`o}}{{Leccia} et~al.}{2007}]{LKB2007}
{Leccia} S.,  {Kjeldsen} H.,  {Bonanno} A.,  {Claudi} R.~U.,  {Ventura} R.,
  {Patern\`o} L.,  2007, A\&A, 464, 1059

\bibitem[\protect\citeauthoryear{Marti{\'c}, {Lebrun}, {Appourchaux} \&
  {Korzennik}}{Marti{\'c} et~al.}{2004}]{MLA2004}
Marti{\'c} M.,  {Lebrun} J.-C.,  {Appourchaux} T.,    {Korzennik} S.~G.,  2004,
  A\&A, 418, 295

\bibitem[\protect\citeauthoryear{{Marti{\'c}}, {Lebrun}, {Appourchaux} \&
  {Schmitt}}{{Marti{\'c}} et~al.}{2004}]{MLA2004b}
{Marti{\'c}} M.,  {Lebrun} J.~C.,  {Appourchaux} T.,    {Schmitt} J.,  2004, in
  Danesy D.,  ed., SOHO 14/GONG 2004 Workshop, Helio- and Asteroseismology:
  Towards a Golden Future ESA SP-559.
p.~563

\bibitem[\protect\citeauthoryear{Matthews, Kuschnig, Guenther et~al.,}{Matthews
  et~al.}{2004}]{MKG2004}
Matthews J.~M.,  Kuschnig R.,  Guenther D.~B.,    et~al., 2004, Nat, 430, 51,
  {Erratum}: 430, 921

\bibitem[\protect\citeauthoryear{{Mosser}, {Bouchy}, {Catala} et~al.,}{{Mosser}
  et~al.}{2005}]{MBC2005}
{Mosser} B.,  {Bouchy} F.,  {Catala} C.,    et~al., 2005, A\&A, 431, L13

\bibitem[\protect\citeauthoryear{North, {Davis}, {Bedding} et~al.,}{North
  et~al.}{2007}]{NDB2007}
North J.~R.,  {Davis} J.,  {Bedding} T.~R.,    et~al., 2007, MNRAS, 380, L83

\bibitem[\protect\citeauthoryear{{R{\'e}gulo} \& {Roca
  Cort{\'e}s}}{{R{\'e}gulo} \& {Roca Cort{\'e}s}}{2005}]{R+RC2005}
{R{\'e}gulo} C.,  {Roca Cort{\'e}s} T.,  2005, A\&A, 444, L5

\bibitem[\protect\citeauthoryear{{Stello}, {Bruntt}, {Kjeldsen}, {Bedding},
  {Arentoft} et~al.,}{{Stello} et~al.}{2007}]{SBK2007}
{Stello} D.,  {Bruntt} H.,  {Kjeldsen} H.,  {Bedding} T.~R.,  {Arentoft} T.,
  et~al., 2007, MNRAS, 377, 584

\bibitem[\protect\citeauthoryear{Stello, Kjeldsen, Bedding et~al.,}{Stello
  et~al.}{2004}]{SKB2004}
Stello D.,  Kjeldsen H.,  Bedding T.~R.,    et~al., 2004, Sol. Phys., 220, 207

\bibitem[\protect\citeauthoryear{{Straka}, {Demarque}, {Guenther}, {Li} \&
  {Robinson}}{{Straka} et~al.}{2006}]{SDG2006}
{Straka} C.~W.,  {Demarque} P.,  {Guenther} D.~B.,  {Li} L.,    {Robinson}
  F.~J.,  2006, ApJ, 636, 1078

\bibitem[\protect\citeauthoryear{{Tarrant}, {Chaplin}, {Elsworth}
  et~al.,}{{Tarrant} et~al.}{2007}]{TCE2007}
{Tarrant} N.~J.,  {Chaplin} W.~J.,  {Elsworth} Y.,    et~al., 2007, MNRAS, in
  press, {\tt\small arXiv:astro-ph/0706.3346}

\end{thebibliography}

\end{document}